\documentclass[twocolumn,showpacs,preprintnumbers,amsmath,amssymb]{revtex4}

\usepackage{graphicx}
\usepackage{dcolumn}
\usepackage{bm}

\begin{document}


\title{Separability criteria for continuous 
variable systems}

\author{Kazuo Fujikawa}
\affiliation{%
Institute of Quantum Science, College of Science and Technology,
Nihon University, Chiyoda-ku, Tokyo 101-8308, Japan
}%


\begin{abstract}
A general separability condition on the second moment (covariance matrix) for continuous variable two-party 
 systems is derived by an analysis analogous to the derivation of the Kennard's uncertainty relation without referring to the non-negativity of the partially transposed density matrix. This
separability criterion is generally more stringent than that used by Simon which is based on the non-negativity of partially transposed density matrix, and thus this criterion may be useful in the analysis of general continuous two-party systems. Another separability criterion used by Duan et al. is shown to be generally weaker than that of Simon. We thus have a hierarchy of separability criteria, but
 all these criteria when combined with suitable squeezing become equivalent at the boundary of the P-representation condition and thus turned out to be sufficient to analyze the separability of two-party Gaussian systems.     
\end{abstract}

\maketitle

\section{Introduction}
The entanglement~\cite{epr} is an intriguing property of 
quantum mechanics, but a general quantitative criterion of 
entanglement appears to be missing at this moment. The negativity or non-negativity of the partially transposed density matrix was proposed as a quantitative means to analyze  the entanglement by Peres \cite{peres}. But this condition is known to work only for simple systems such as a two-spin system~\cite{horodecki}. In view of this situation, it is 
remarkable that the necessary and sufficient conditions exist for the separability of two-party Gaussian systems~\cite{simon,duan}. 
 To be precise, the "two-party systems" in the present paper are used for the systems with one freedom (or one-mode) in each party, since it is known that the analysis of the separability criterion for the two-party systems with more than one freedom in each party is more involved \cite{werner}.
The criterion used by Simon~\cite{simon} is based on a generalization of the non-negativity condition of the partial transposed density matrix in the manner of Peres.
On the other hand, the criterion used by Duan et al.~\cite{duan} is based on a set of relations for EPR-type operators. The non-negativity condition was further analyzed by 
Shchukin and Vogel \cite{shchukin, miranowicz}. 
This separability and related issues have been discussed in 
the past by 
various authors, for example, in~\cite{englert, giedke, mancini, eisert,
vidal, wolf, raymer, giovannetti, giedke2}.
The present status of the quantum separability problem of 
two-party Gaussian states is  reviewed in~\cite{mancini2}.
See also a review of continuous variable systems in \cite{braunstein}.

We here start with an elementary analysis of the Heisenberg uncertainty relation in the manner of 
Kennard~\cite{kennard, robertson} and we derive a separability condition, which is generally more stringent than the separability condition given by Simon~\cite{simon}, {\em without} referring to the negativity of the partially transposed density matrix. This implies that our criterion is more efficient as a condition for separability than the condition used by Simon, which is based on the non-negativity of the partially transposed density matrix, although neither criterion provides the sufficient separability condition for two-party systems in general. Our analysis is complementary to that of Shchukin and Vogel \cite{shchukin, miranowicz}; they analyze higher moments also in the framework of the non-negativity of the partially transposed density matrix, while we specialize in the second moment and search for a more stringent condition. We also show that another class of separability conditions used by Duan et al. \cite{duan} are generally weaker than the condition of Simon.
Despite these differences, all these criterions when combined with suitable squeezing give rise to the 
necessary and sufficient condition for the separability of two-party Gaussian systems. This fact has been shown elsewhere
by using the analytic solutions of squeezing parameters \cite{fujikawa} which establish the equivalence of the P-representation condition for Gaussian systems with the $Sp(2,R)\otimes Sp(2,R)$ invariant separability condition.
Some of the technical aspects involved in the analysis in \cite{fujikawa} are also clarified in the present paper.

\section{Entanglement and Kennard's relation}

\subsection{Kennard's relation}

We consider a two-party
system with one freedom in each party (or a two-particle system in one-dimensional space) 
described by canonical variables $(q_{1}, p_{1})$ and 
$(q_{2}, p_{2})$.
We define 
\begin{eqnarray}
&&\hat{X}(d,f)=d_{1}\hat{q}_{1}+d_{2}\hat{p}_{1}
+f_{1}\hat{q}_{2}+f_{2}\hat{p}_{2},\nonumber\\ 
&&\hat{X}(g,h)=g_{1}\hat{q}_{1}+g_{2}\hat{p}_{1}
+h_{1}\hat{q}_{2}+h_{2}\hat{p}_{2}
\end{eqnarray}
where all the coefficients
\begin{eqnarray}
&&d^{T}=(d_{1},d_{2}), \ f^{T}=(f_{1},f_{2}), 
\ g^{T}=(g_{1},g_{2}),\nonumber\\
&&h^{T}=(h_{1},h_{2})
\end{eqnarray}
are real numbers. The use of the operators in (1) is known to be 
convenient to define the general uncertainty relations. In fact, the conventional form of the Kennard's relation is derived as a 
special case, as is shown in (14) below.
We deal with 
a mixed state $\hat{\rho}=\sum_{k}P_{k}|\psi_{k}\rangle\langle
\psi_{k}|$ with $P_{k}\geq 0$ and $\sum_{k}P_{k}=1$. 

The Kennard's relation for a specific {\em pure} state $|\psi_{k}\rangle$
is written for any choice of parameters $d$ to $h$ as 
\begin{eqnarray}
&&\langle(\hat{X}(d,f)-\langle\hat{X}(d,f)\rangle_{k})^{2}\rangle_{k}+
\langle(\hat{X}(g,h)-\langle\hat{X}(g,h)\rangle_{k})^{2}\rangle_{k}\nonumber\\
&&\geq |d^{T}Jg+f^{T}Jh|
\end{eqnarray}
where the $2\times 2$ matrix is given by
\begin{eqnarray}
J=\left(\begin{array}{cc}
  0&1\\
  -1&0\\
            \end{array}\right),
\end{eqnarray} 
and $\langle\hat{X}(d,f)\rangle_{k}=\langle\psi_{k}|\hat{X}(d,f)|\psi_{k}\rangle$, for example.
The relation (3) is derived from 
$\langle
\psi_{k}|\hat{\eta}\hat{\eta}^{\dagger}|\psi_{k}\rangle\geq0$ and 
$\langle\psi_{k}|\hat{\eta}^{\dagger}\hat{\eta}|\psi_{k}\rangle\geq0$ for
\begin{eqnarray}
\hat{\eta}=\hat{X}(d,f)-\langle\hat{X}(d,f)\rangle_{k} +
i(\hat{X}(g,h)-\langle\hat{X}(g,h)\rangle_{k}),
\end{eqnarray}
and the right-hand side of (3) stands for the commutator 
\begin{eqnarray}
&&|[\hat{X}(d,f)-\langle\hat{X}(d,f)\rangle_{k},
\hat{X}(g,h)-\langle\hat{X}(g,h)\rangle_{k}]|\nonumber\\
&&=|d^{T}Jg+f^{T}Jh|.
\nonumber
\end{eqnarray} 
We thus conclude by taking the weighted sum of (3)
\begin{eqnarray}
&&\sum_{k}P_{k}\langle(\hat{X}(d,f)-\langle\hat{X}(d,f)\rangle_{k})^{2}\rangle_{k}\nonumber\\
&&+ \sum_{k}P_{k}
\langle(\hat{X}(g,h)-\langle\hat{X}(g,h)\rangle_{k})^{2}\rangle_{k}\nonumber\\
&&\geq |d^{T}Jg+f^{T}Jh|
\end{eqnarray}

On the other hand, it is customary to deal with 
\begin{eqnarray}
\langle(\Delta\hat{X}(d,f))^{2}\rangle=
Tr\{(\Delta\hat{X}(d,f))^{2}\hat{\rho}\}
\end{eqnarray}
where
$\Delta\hat{X}(d,f)=\hat{X}(d,f)-\langle\hat{X}(d,f)\rangle$
with
\begin{eqnarray}
\langle\hat{X}(d,f)\rangle=Tr\{\hat{X}(d,f)\hat{\rho}\}
\end{eqnarray}
in the analysis of uncertainty relations for a mixed state.
We note the relation
\begin{eqnarray}
\langle(\Delta\hat{X}(d,f))^{2}\rangle
&=&\langle\left(\hat{X}(d,f)-\langle\hat{X}(d,f)\rangle
\right)^{2}\rangle
\nonumber\\
&=&\sum_{k}P_{k}\langle\left(\hat{X}(d,f)
-\langle\hat{X}(d,f)\rangle\right)^{2}
\rangle_{k}\nonumber\\
&=&\sum_{k}P_{k}\langle [ \hat{X}(d,f)
-\langle\hat{X}(d,f)\rangle_{k}\nonumber\\
&& +\langle\hat{X}(d,f)\rangle_{k}
-\langle\hat{X}(d,f)\rangle ]^{2}
\rangle_{k}\nonumber\\
&=&\sum_{k}P_{k}[\langle\left(\hat{X}(d,f)
-\langle\hat{X}(d,f)\rangle_{k}
\right)^{2}\rangle_{k}\nonumber\\
&&+\left(\langle\hat{X}(d,f)\rangle_{k}-
\langle\hat{X}(d,f)\rangle\right)^{2}].
\end{eqnarray}
From (6) and (9) we thus conclude the basic relation 
\begin{eqnarray}
&&\langle(\Delta\hat{X}(d,f))^{2}\rangle+\langle(\Delta\hat{X}(g,h))^{2}\rangle\nonumber\\
&&=\sum_{k}P_{k}[\langle\left(\hat{X}(d,f)
-\langle\hat{X}(d,f)\rangle_{k}
\right)^{2}\rangle_{k}\nonumber\\
&&+\left(\langle\hat{X}(d,f)\rangle_{k}-
\langle\hat{X}(d,f)\rangle\right)^{2}]\nonumber\\
&&+\sum_{k}P_{k}[\langle\left(\hat{X}(g,h)
-\langle\hat{X}(g,h)\rangle_{k}
\right)^{2}\rangle_{k}\nonumber\\
&&+\left(\langle\hat{X}(g,h)\rangle_{k}-
\langle\hat{X}(g,h)\rangle\right)^{2}]\nonumber\\
&&\geq  
\sum_{k}P_{k}\left(\langle\hat{X}(d,f)\rangle_{k}-
\langle\hat{X}(d,f)\rangle\right)^{2}\nonumber\\
&&+\sum_{k}P_{k}\left(\langle\hat{X}(g,h)\rangle_{k}-
\langle\hat{X}(g,h)\rangle\right)^{2}\nonumber\\
&&+|d^{T}Jg+f^{T}Jh|.
\end{eqnarray}
This relation  is more precise than the customary form of the uncertainty 
relation for a mixed state
\begin{eqnarray}
\langle(\Delta\hat{X}(d,f))^{2}\rangle+\langle(\Delta\hat{X}(g,h))^{2}\rangle\geq |d^{T}Jg+f^{T}Jh|
\end{eqnarray}
which is obtained from 
$Tr\{\hat{\eta}^{\prime}(\hat{\eta}^{\prime})^{\dagger}
\hat{\rho}\}\geq0$ and 
$Tr\{(\hat{\eta}^{\prime})^{\dagger}\hat{\eta}^{\prime}\hat{\rho}\}\geq0$ with
\begin{eqnarray}
\hat{\eta}^{\prime}=\Delta\hat{X}(d,f)+i\Delta\hat{X}(g,h),
\nonumber
\end{eqnarray}
and the right-hand side of (11) standing for the commutator $|[\Delta\hat{X}(d,f),
\Delta\hat{X}(g,h)]|$. 
Note that one may choose 
$\langle\hat{X}(d,f)\rangle
=\sum_{k}P_{k}\langle\hat{X}(d,f)\rangle_{k}=0$, but $\langle\hat{X}(d,f)\rangle_{k}$ for each component state with
\begin{eqnarray} 
\langle\hat{X}(d,f)\rangle_{k}
=\int dq_{1}dq_{2}\psi^{\star}_{k}(q_{1},q_{2})\hat{X}(d,f)
\psi_{k}(q_{1},q_{2})
\end{eqnarray}
does not vanish in general in (10).

In passing, the Kennard's relation for a general pure state (3) is also written as 
\begin{eqnarray}
&&x^{2}\langle\left(\hat{X}(d,f)
-\langle\hat{X}(d,f)\rangle_{k}\right)^{2}\rangle_{k}\\
&&+\langle\left(\hat{X}(g,h)
-\langle\hat{X}(g,h)\rangle_{k}\right)^{2}
\rangle_{k}\geq 
x|d^{T}Jg+f^{T}Jh|\nonumber
\end{eqnarray}
by replacing $d\rightarrow xd, \ f\rightarrow xf$ for any real
positive $x$ (and this relation holds for negative $x$ also), and thus the discriminant gives the conventional form of
Kennard's relation
\begin{eqnarray}
&&\sqrt{\langle\left(\hat{X}(d,f)
-\langle\hat{X}(d,f)\rangle_{k}\right)^{2}
\rangle_{k}}\sqrt{\langle\left(\hat{X}(g,h)
-\langle\hat{X}(g,h)\rangle_{k}\right)^{2}
\rangle_{k}}\nonumber\\
&&\geq \frac{1}{2}|d^{T}Jg+f^{T}Jh|
\end{eqnarray}
if the parameters $d$ to $h$ are suitably chosen.

\subsection{Separability condition}

For a separable pure state 
$\psi_{k}(q_{1},q_{2})=\phi_{k}(q_{1})\varphi_{k}(q_{2})$, 
we have
\begin{eqnarray}
&&\langle\left(\hat{X}(d,f)
-\langle\hat{X}(d,f)\rangle_{k}\right)^{2}
\rangle_{k}\nonumber\\
&&=\langle\left(d_{1}\hat{q}_{1}
+d_{2}\hat{p}_{1}
-\langle d_{1}\hat{q}_{1}+d_{2}\hat{p}_{1}\rangle_{k}\right)
^{2}\rangle_{k}\nonumber\\
&&+\langle\left(f_{1}\hat{q}_{2}
+f_{2}\hat{p}_{2}
-\langle f_{1}\hat{q}_{2}+f_{2}\hat{p}_{2}
\rangle_{k}\right)^{2}\rangle_{k}
\end{eqnarray}
since the cross terms vanish,
and similarly for $\langle\left(\hat{X}(g,h)
-\langle\hat{X}(g,h)\rangle_{k}\right)^{2}
\rangle_{k}$.
We  thus have for a separable pure state
\begin{eqnarray}
&&\langle\left(\hat{X}(d,f)
-\langle\hat{X}(d,f)\rangle_{k}\right)^{2}
\rangle_{k}+\langle\left(\hat{X}(g,h)
-\langle\hat{X}(g,h)\rangle_{k}\right)^{2}
\rangle_{k}\nonumber\\
&&=\langle\left(\hat{X}(d,0)
-\langle\hat{X}(d,0)\rangle_{k}\right)^{2}
\rangle_{k}\nonumber\\
&&+\langle\left(\hat{X}(g,0)
-\langle\hat{X}(g,0)\rangle_{k}\right)^{2}
\rangle_{k}\nonumber\\
&&+
\langle\left(\hat{X}(0,f)
-\langle\hat{X}(0,f)\rangle_{k}\right)^{2}
\rangle_{k}\nonumber\\
&&+\langle\left(\hat{X}(0,h)
-\langle\hat{X}(0,h)\rangle_{k}\right)^{2}
\rangle_{k}\nonumber\\
&&\geq|d^{T}Jg| +
|f^{T}Jh|
\end{eqnarray}
which holds for any choice of parameters $d$ to $h$.  Here we used (3)
for $f=h=0$ or $d=g=0$.
The equality sign holds only for 
\begin{eqnarray}
&&[\left(d_{1}\hat{q}_{1}
+d_{2}\hat{p}_{1}
-\langle d_{1}\hat{q}_{1}+d_{2}\hat{p}_{1}\rangle_{k}\right)
\nonumber\\
&&+i\left(g_{1}\hat{q}_{1}
+g_{2}\hat{p}_{1}
-\langle g_{1}\hat{q}_{1}+g_{2}\hat{p}_{1}
\rangle_{k}\right)]\phi_{k}(q_{1})=0,
\nonumber\\
&&[\left(f_{1}\hat{q}_{2}
+f_{2}\hat{p}_{2}
-\langle f_{1}\hat{q}_{2}+f_{2}\hat{p}_{2}\rangle_{k}\right)
\nonumber\\
&&+i\left(h_{1}\hat{q}_{2}
+h_{2}\hat{p}_{2}
-\langle h_{1}\hat{q}_{2}+ h_{2}\hat{p}_{2}
\rangle_{k}\right)]\varphi_{k}(q_{2})=0, \nonumber
\end{eqnarray}
for a suitable choice of parameters $d$ to $h$.

We finally conclude from the first expression in (10) when combined with (16) for any separable density matrix
\begin{eqnarray}
&&\langle\left(\Delta\hat{X}(d,f)\right)^{2}\rangle+
\langle\left(\Delta\hat{X}(g,h)\right)^{2}\rangle
\nonumber\\
&&\geq
\sum_{k}P_{k}[\left(\langle\hat{X}(d,f)\rangle_{k}-
\langle\hat{X}(d,f)\rangle\right)^{2}\nonumber\\
&&\hspace{1 cm} +\left(\langle\hat{X}(g,h)\rangle_{k}-
\langle\hat{X}(g,h)\rangle\right)^{2}]\nonumber\\
&&+|d^{T}Jg| +|f^{T}Jh|
\end{eqnarray}
which holds for any choice of parameters $d$ to $h$. This relation (17) is our basic {\em necessary condition} for the separability of general two-party systems.
 
We next define the variables
$(\hat{\xi}_{\mu})=(\hat{q}_{1}, \hat{p}_{1}, \hat{q}_{2}, 
\hat{p}_{2})$ and the $4\times 4$ correlation matrix $V$ by
\begin{eqnarray}
&&V=(V_{\mu\nu}),\\   
&&V_{\mu\nu}=\frac{1}{2}
\langle\Delta\hat{\xi}_{\mu}\Delta\hat{\xi}_{\nu}+
\Delta\hat{\xi}_{\nu}\Delta\hat{\xi}_{\mu}\rangle
=\frac{1}{2}
\langle\{\Delta\hat{\xi}_{\mu}, \Delta\hat{\xi}_{\nu}\}
\rangle\nonumber
\end{eqnarray}
where $\Delta\hat{\xi}_{\mu}=\hat{\xi}_{\mu}
-\langle\hat{\xi}_{\mu}\rangle$ with $\langle\hat{\xi}_{\mu}\rangle=\sum_{k}P_{k}\langle
\psi_{k}|\hat{\xi}_{\mu}|\psi_{k}\rangle$. 
 We here recall the definition of the density matrix $\hat{\rho}=\sum_{k}P_{k}|\psi_{k}\rangle\langle
\psi_{k}|$ with $P_{k}\geq 0$ and $\sum_{k}P_{k}=1$.
The correlation matrix  can be written in the form 
\begin{eqnarray}
V=\left(\begin{array}{cc}
  A&C\\
  C^{T}&B\\
            \end{array}\right)
\end{eqnarray}
where $A$ and $B$ are $2\times 2$ real symmetric matrices and 
$C$ is a $2\times 2$ real matrix. We also define 
\begin{eqnarray}
\tilde{V}=(\tilde{V}_{\mu\nu}), \ \ \  
\tilde{V}_{\mu\nu}=\sum_{k}P_{k}
\langle\Delta\hat{\xi}_{\mu}\rangle_{k}
\langle\Delta\hat{\xi}_{\nu}\rangle_{k}
\end{eqnarray}
and
\begin{eqnarray}
\tilde{V}=\left(\begin{array}{cc}
  \tilde{A}&\tilde{C}\\
  \tilde{C}^{T}&\tilde{B}\\
            \end{array}\right)
\end{eqnarray}
where $\tilde{A}$ and $\tilde{B}$ are $2\times 2$ real symmetric
 matrices and 
$\tilde{C}$ is a $2\times 2$ real matrix. Both of $V$ and 
$\tilde{V}$ are non-negative.

The basic relation (17) for separable states, which is a necessary condition for separability, is then written as
\begin{eqnarray}
&&d^{T}Ad+f^{T}Bf+2d^{T}Cf+g^{T}Ag+h^{T}Bh+2g^{T}Ch\nonumber\\
&&\geq d^{T}\tilde{A}d+f^{T}\tilde{B}f+2d^{T}\tilde{C}f
+g^{T}\tilde{A}g+h^{T}\tilde{B}h+2g^{T}\tilde{C}h
\nonumber\\
&&+|d^{T}Jg| + |f^{T}Jh|.
\end{eqnarray}
while the  Kennard's relation for general states (10) is written as 
\begin{eqnarray}
&&d^{T}Ad+f^{T}Bf+2d^{T}Cf+g^{T}Ag+h^{T}Bh+2g^{T}Ch\nonumber\\
&&\geq d^{T}\tilde{A}d+f^{T}\tilde{B}f+2d^{T}\tilde{C}f
+g^{T}\tilde{A}g+h^{T}\tilde{B}h+2g^{T}\tilde{C}h
\nonumber\\
&&+|d^{T}Jg+f^{T}Jh|.
\end{eqnarray}
Note the difference between $|d^{T}Jg| + |f^{T}Jh|$ and 
$|d^{T}Jg+f^{T}Jh|$. The appearance of the non-negative matrix $\tilde{V}$ on the right-hand sides of these relations is the novel aspect
of our formulation.
The anti-symmetric commutator parts in 
\begin{eqnarray}
\langle\Delta\hat{\xi}_{\mu}\Delta\hat{\xi}_{\nu}\rangle
=\frac{1}{2}
\langle\{\Delta\hat{\xi}_{\mu}, \Delta\hat{\xi}_{\nu}\}
\rangle
+\frac{1}{2}
\langle[\Delta\hat{\xi}_{\mu}, \Delta\hat{\xi}_{\nu}]\rangle,
\end{eqnarray}
which may be added to $A$ and $B$ in (19), do not contribute 
to (22) since the parameters $d$ to $h$ are all real.

Under the 
$S_{1}\otimes S_{2}\in Sp(2,R)\otimes Sp(2,R)$  transformations 
of
$(\hat{q}_{1},\hat{p}_{1})$ and $(\hat{q}_{2},\hat{p}_{2})$,
respectively, we have
\begin{eqnarray}
&&A\rightarrow S_{1}AS^{T}_{1},\ \ \ 
B\rightarrow S_{2}BS^{T}_{2},\ \ \ C\rightarrow S_{1}CS^{T}_{2}
\nonumber\\
&&\tilde{A}\rightarrow S_{1}\tilde{A}S^{T}_{1},\ \ \ 
\tilde{B}\rightarrow S_{2}\tilde{B}S^{T}_{2},\ \ \ \tilde{C}
\rightarrow S_{1}\tilde{C}S^{T}_{2}
\end{eqnarray}
which is equivalent to the transformation 
\begin{eqnarray}
d\rightarrow S^{T}_{1}d,\ \ \ 
f\rightarrow S^{T}_{2}f,\ \ \ g\rightarrow S^{T}_{1}g,
 \ \ \ h\rightarrow S^{T}_{2}h
\end{eqnarray}
in (22) if one recalls $J=S_{1}JS^{T}_{1}$ and $J=S_{2}JS^{T}_{2}$
; the inequality (22), which is valid for {\em any} choice of parameters $d$ to $h$, holds
 after the transformation (26) and in this sense (22) is 
invariant under the above $Sp(2,R)\otimes Sp(2,R)$. To be 
precise, we do not use any property of the wave function under 
$Sp(2,R)\otimes Sp(2,R)$, and thus our $Sp(2,R)\otimes Sp(2,R)$ 
transformation is rather defined by (25) for given constant
matrices $A,\ B$ and $C$.

The difference between the two expressions in (22) and 
(23) appears when one 
replaces $B$ and $C$ by $S^{T}_{3}BS_{3}$ and $CS_{3}$ (and 
also $\tilde{B}$ and $\tilde{C}$ by 
$S^{T}_{3}\tilde{B}S_{3}$ and $\tilde{C}S_{3}$), respectively,  
with 
\begin{eqnarray}
S_{3}=\left(\begin{array}{cc}
  1&0\\
  0&-1\\
            \end{array}\right).
\end{eqnarray}
One can undo the replacements  in 
the expression (22) by transformations 
$f\rightarrow S_{3}f$ and $h\rightarrow S_{3}h$, whereas it 
leads to $|d^{T}Jg-f^{T}Jh|$ in the relation 
(23). The separability condition thus demands that the Kennard's relation, when written in the form (22), should hold both for the original system and for the system with the replacements of $B$ and $C$ by $S^{T}_{3}BS_{3}$ and $CS_{3}$, respectively, which may {\em a priori} correspond to an  unphysical density matrix for {\em inseparable} systems. By using $S_{3}$, one can adjust the signature of ${\rm det}C$ at one's will~\cite{simon} in the separability condition (22).

It is also useful to consider the separability conditions {\em weaker} than the separability condition (22) by imposing 
{\em subsidiary conditions} $g=J^{T}d$ and 
$h=J^{T}f$ or $g=J^{T}d$ and $h=-J^{T}f$, respectively,
\begin{eqnarray}
&&d^{T}Ad+f^{T}Bf+2d^{T}Cf+d^{T}JAJ^{T}d\nonumber\\
&&+f^{T}JBJ^{T}f
\pm 2d^{T}JCJ^{T}f\nonumber\\
&&\geq d^{T}\tilde{A}d+f^{T}\tilde{B}f
+2d^{T}\tilde{C}f+d^{T}J\tilde{A}J^{T}d
\nonumber\\
&&+f^{T}J\tilde{B}J^{T}f
\pm 2d^{T}J\tilde{C}J^{T}f
\nonumber\\
&&+(d^{T}d+f^{T}f)
\end{eqnarray}
which is also written as 
\begin{eqnarray}
&&\left(\begin{array}{cc}
  A+JAJ^{T}&C\pm JCJ^{T}\\
  C^{T}\pm JC^{T}J^{T}&B+JBJ^{T}\\
            \end{array}\right)
\nonumber\\
&&\geq 
\left(\begin{array}{cc}
  \tilde{A}+J\tilde{A}J^{T}&\tilde{C}\pm J\tilde{C}J^{T}\\
  \tilde{C}^{T}\pm J\tilde{C}^{T}J^{T}&\tilde{B}+J\tilde{B}J^{T}\\
            \end{array}\right) + I.
\end{eqnarray}
For general inseparable  states in (23), we 
have only the first condition in (28) if one wants to keep 
$(d^{T}d+f^{T}f)$ on the right-hand side in this form. The 
basic  
$Sp(2,R)\otimes Sp(2,R)$ invariance of uncertainty relations 
(22) and (23)
is lost in these weaker conditions (28) and (29) with subsidiary conditions, 
but they have applications in the analysis of the 
P-representation. See Appendix B for the definition of 
the P-representation of Gaussian states.

\section{Separability and P-representation of Gaussian states}

\subsection{General analysis}

One can bring any given $V$ in (19) by $Sp(2,R)\otimes Sp(2,R)$
transformations to the standard form~\cite{ simon, duan} (see
also Appendix A)
\begin{eqnarray}
V_{0}=\left(\begin{array}{cccc}
  a&0&c_{1}&0\\
  0&a&0&c_{2}\\
  c_{1}&0&b&0\\
  0&c_{2}&0&b\\            
  \end{array}\right).
\end{eqnarray}

One may understand (22) and (23) ( and similarly
(28) and (29)), which hold for any choice of parameters $d$ to 
$h$, as imposing constraints on the allowed
ranges  of the elements $a, b, c_{1}, c_{2}$ of the standard 
form of $V_{0}$ in (30), for example. This  
interpretation was also adopted by Simon~\cite{simon}. 
In this interpretation, the full relation (22)
is more restrictive than the  relations (28) and (29) with 
the {\em subsidiary conditions} on the parameters $d$ to $h$. 
In other words, the elements $a, b, c_{1}, c_{2}$ which satisfy
(22) automatically satisfy (28) and (29), but not the 
other way around. In our analysis below, we interpret these 
relations as constraints on $c_{1}, c_{2}$ for fixed $a,b$.  \footnote{Alternatively, one may understand the relations (22) and (23) as an infinite set of uncertainty 
relations (and their variants) for each given set of parameters $d$ to $h$. } 
 
We now analyze the Gaussian states.
The existence of the well-defined P-representation of two-party Gaussian states requires the condition (see Appendix B)
\begin{eqnarray}
P^{-1}= V - \frac{1}{2}I \geq 0
\end{eqnarray}
namely
\begin{eqnarray}
d^{T}Ad+f^{T}Bf+2d^{T}Cf\geq \frac{1}{2}(d^{T}d+f^{T}f)
\end{eqnarray}
for any $d$ and $f$. 
By using a special property of the P-representation, namely,
a special property of the coherent state, one can identify
 $\tilde{V}$ in (21) with the matrix $P^{-1}$ if the P-representation exists. See (B13) in Appendix B.
We thus have
\begin{eqnarray}
P^{-1}=\tilde{V}=V-\frac{1}{2}I
\end{eqnarray}
or equivalently
\begin{eqnarray}
&&d^{T}\tilde{A}d+f^{T}\tilde{B}f
+2d^{T}\tilde{C}f\nonumber\\
&&=d^{T}Ad+f^{T}Bf+2d^{T}Cf
-\frac{1}{2}(d^{T}d+f^{T}f)
\end{eqnarray}
which is in fact non-negative for any $d$ and $f$, as is required by (32), since $\tilde{V}$ is positive semi-definite by its construction. 

We first show that the P-representation condition implies the 
separability conditions. When one adds (34) to the expression obtained from (34) by the replacement of $d$ and $f$ by 
$g$ and $h$, respectively, 
one reproduces the separability condition (22)
\begin{eqnarray}
&&d^{T}Ad+f^{T}Bf+2d^{T}Cf+g^{T}Ag+h^{T}Bh+2g^{T}Ch\nonumber\\
&&= d^{T}\tilde{A}d+f^{T}\tilde{B}f+2d^{T}\tilde{C}f
+g^{T}\tilde{A}g+h^{T}\tilde{B}h+2g^{T}\tilde{C}h
\nonumber\\
&&+\frac{1}{2}(d^{T}d+f^{T}f)+\frac{1}{2} (g^{T}g+h^{T}h)
\nonumber\\
&&\geq d^{T}\tilde{A}d+f^{T}\tilde{B}f+2d^{T}\tilde{C}f
+g^{T}\tilde{A}g+h^{T}\tilde{B}h+2g^{T}\tilde{C}h
\nonumber\\
&&+ |d^{T}Jg|+|f^{T}Jh|,
\end{eqnarray}
where we used
\begin{eqnarray}
&&\frac{1}{2}(d^{T}d+g^{T}g)+\frac{1}{2} (f^{T}f+h^{T}h)
\nonumber\\
&&\geq \sqrt{(d^{T}d)(g^{T}J^{T}Jg)}+\sqrt{ (f^{T}f)(h^{T}J^{T}Jh)}\nonumber\\
&&\geq
|d^{T}Jg|+|f^{T}Jh|.
\end{eqnarray}
This is 
natural since the P-representation of Gaussian states is in fact separable. 

But the converse is not obvious and needs to be proved. The separability condition 
(22) is invariant under $Sp(2,R)\otimes Sp(2,R)$ in (25), 
whereas
the P-representation condition  (31) or (32) is not invariant under 
$S_{1}\otimes S_{2}\in Sp(2,R)\otimes Sp(2,R)$ by  noting that 
\begin{eqnarray}
d^{T}S_{1}^{T}S_{1}d+f^{T}S_{2}^{T}S_{2}f \neq 
d^{T}d+f^{T}f
\end{eqnarray}
in general. In this sense these two 
conditions cannot be equivalent to each other.
In the process of the identification (33), the $Sp(2,R)\otimes Sp(2,R)$ invariance of $\tilde{V}$ is lost due to the normal ordering operation in the P-representation. See Appendix B.

The separability condition of Simon \cite{simon}, which is based 
on the non-negativity of the partially transposed density matrix, is given by 
\begin{eqnarray}
V+\frac{i}{2}\left(\begin{array}{cc}
  J&0\\
  0&\pm J\\            
  \end{array}\right)\geq 0
\end{eqnarray}
or equivalently (by taking the average of (38) in the form $v^{\dagger}M v$ by the four-component
complex vectors $v=(d\pm ig, f\pm ih)$)
\begin{eqnarray}
&&d^{T}Ad+f^{T}Bf+2d^{T}Cf+g^{T}Ag+h^{T}Bh+2g^{T}Ch
\nonumber\\
&&\geq |d^{T}Jg|+|f^{T}Jh|
\end{eqnarray}
which is $Sp(2,R)\otimes Sp(2,R)$ invariant.
Since $\tilde{V}$ is non-negative, the condition (22) is generally  more stringent than (39), and thus the condition (38)
or (39) is necessary but {\em not sufficient} for separability in 
general even for the case when (22) is sufficient. However, for the P-representation of Gaussian states, we have 
 $\tilde{V}=P^{-1}=V-(1/2)I$ as in (33) and thus $\tilde{V}$ is determined by $V$. Moreover at the boundary of the P-representation condition, the two eigenvalues of $V-(1/2)I$ after a suitable squeezing transformation vanish \cite{fujikawa}. For this reason, the separability condition (39) can be equivalent to (22) and thus can also be sufficient for the separability of Gaussian states. This is indeed the case as we have shown elsewhere by an explicit algebraic analysis \cite{fujikawa}.
  
  There is yet another set of separability conditions formulated by Duan et al. \cite{duan}. To be precise, the condition used in \cite{duan} corresponds to the {\em weaker} separability condition (29) 
\begin{eqnarray}
\left(\begin{array}{cc}
  A+JAJ^{T}&C\pm JCJ^{T}\\
  C^{T}\pm JC^{T}J^{T}&B+JBJ^{T}\\
            \end{array}\right)
\geq I
\end{eqnarray} 
applied to the separability condition of Simon (39) by setting $\tilde{V}=0$. This fact, which has been briefly mentioned elsewhere \cite{fujikawa}, is explained in more detail in the following.

\subsection{Quantitative analysis}

We start with the analysis of the P-representation condition 
$V-(1/2)I\geq 0$ in (31). We apply this condition to the 
covariance matrix 
\begin{eqnarray}
V&=&S^{-1}V_{0}(S^{-1})^{T}\nonumber\\
&=&\left(\begin{array}{cccc}
  ar_{1}&0&c_{1}\sqrt{r_{1}r_{2}}&0\\
  0&a/r_{1}&0&c_{2}/\sqrt{r_{1}r_{2}}\\
  c_{1}\sqrt{r_{1}r_{2}}&0&br_{2}&0\\
  0&c_{2}/\sqrt{r_{1}r_{2}}&0&b/r_{2}\\            
  \end{array}\right)
\end{eqnarray}
which is obtained from the standard form $V_{0}$ in (30) by
a transformation 
$S(r_{1},r_{2})\in Sp(2,R)\otimes Sp(2,R)$ defined by 
\begin{eqnarray}
S(r_{1},r_{2})S^{T}(r_{1},r_{2})=\left(\begin{array}{cccc}
  1/r_{1}&0&0&0\\
  0&r_{1}&0&0\\
  0&0&1/r_{2}&0\\
  0&0&0&r_{2}\\            
\end{array}\right)
\end{eqnarray}
with suitably chosen $r_{1}\geq 1$ and $r_{2}\geq 1$.
By evaluating eigenvalues of (41), one then obtains the conditions for $V-(1/2)I\geq 0$ \cite{fujikawa}
\begin{eqnarray}
&&(a-\frac{1}{2r_{1}})(b-\frac{1}{2r_{2}})\geq 
c_{1}^{2},
\nonumber\\
&&
(a-\frac{1}{2}r_{1})(b-\frac{1}{2}r_{2})\geq c_{2}^{2}
\end{eqnarray}
together with
\begin{eqnarray} 
(a-\frac{1}{2r_{1}})+(b-\frac{1}{2r_{2}})\geq 0, \ \ \ 
(a-\frac{1}{2}r_{1})+(b-\frac{1}{2}r_{2})\geq 0
\end{eqnarray}
which require $a\geq 1/2$ and $b\geq 1/2$.
The boundary of the P-representation conditions (43), namely, the extremal of these conditions with respect to $r_{1}$ and $r_{2}$ when regarded as the bound on 
$c^{2}_{1}$ is specified by
\begin{eqnarray}
(a-\frac{1}{2r_{1}})(b-\frac{1}{2r_{2}})
&=&\frac{1}{t^{2}}
[(a-\frac{1}{2}r_{1})(b-\frac{1}{2}r_{2})]
\end{eqnarray}
together with the constraint which defines the extremality  \cite{fujikawa}
\begin{eqnarray}
\frac{(ar_{1}-1/2)}{(a/r_{1}-1/2)}=
\frac{(br_{2}-1/2)}{(b/r_{2}-1/2)}.
\end{eqnarray}
Here we defined 
\begin{eqnarray}
0 \leq t\equiv |c_{2}|/|c_{1}|\leq 1
\end{eqnarray}
by choosing $|c_{2}|\leq |c_{1}|$ without loss of generality.
The conditions (45) and (46) are explicitly solved as \cite{fujikawa}
\begin{eqnarray}
&&r_{1}=\frac{1}{at+b}\{ab(1-t^{2})+
\sqrt{D(a,b,t)}\},\nonumber\\
&&r_{2}=\frac{1}{a+bt}\{ab(1-t^{2})+
\sqrt{D(a,b,t)}\}
\end{eqnarray}
with the auxiliary quantity 
\begin{eqnarray}
D(a,b,t)=\sqrt{a^{2}b^{2}(1-t^{2})^{2}+t(a+bt)(at+b)}.
\end{eqnarray}
These squeezing parameters are shown to satisfy \cite{fujikawa}
\begin{eqnarray}
2a\geq r_{1}\geq 1, \ \ \ \ 2b\geq r_{2}\geq 1
\end{eqnarray}
for $a\geq \frac{1}{2}$ and $b\geq \frac{1}{2}$, to be consistent with (44).

For these values of squeezing parameters in (48), we have the upper bound on $|c_{1}|$ from (43) by considering the square root of (45) \cite{fujikawa}
\begin{eqnarray}
|c_{1}|&\leq&\frac{\sqrt{(ar_{1}-1/2)(br_{2}-1/2)}}{\sqrt{r_{1}r_{2}}}\nonumber\\
&&=\frac{ \sqrt{(a/r_{1}-1/2)(b/r_{2}-1/2)}}{(t/\sqrt{r_{1}r_{2}})}
\nonumber\\
&&=\frac{1}{2t}\{[2ab(1+t^{2})+t]-2\sqrt{D(a,b,t)}\}^{1/2}
\end{eqnarray}
The condition (51) defines the P-representation condition which characterizes the separable Gaussian states, namely, we have the P-representation for any $|c_{1}|$ which satisfies (51) for any given $a\geq \frac{1}{2},\ b\geq \frac{1}{2}$ and $1\geq t\geq 0$.
 
The separability condition (38) or (39) derived by Simon is
written as the algebraic conditions
\begin{eqnarray}
&&4(ab-c_{1}^{2})(ab-c_{2}^{2})\geq (a^{2}+b^{2})+2|c_{1}c_{2}|
-\frac{1}{4},\nonumber\\
&&\sqrt{(2a-1)(2b-1)}\geq |c_{1}|+|c_{2}|
\end{eqnarray}
together with $a\geq 1/2$ and $b\geq 1/2$. The second condition in (52) is derived from the weaker condition (40) applied to the standard form $V_{0}$ in (30) \cite{fujikawa}; the second condition is used to exclude the solutions with $|c_{1}|\rightarrow \infty$ and $|c_{1}|\rightarrow \infty$ for fixed $a$ and $b$, which are allowed by the first condition in (52) of Simon. These conditions (52) 
are explicitly solved as \cite{fujikawa}
\begin{eqnarray}
c_{1}^{2}
&\leq&\frac{1}{4t^{2}}\{[2ab(1+t^{2})+t]-2\sqrt{D(a,b,t)}\},\nonumber\\
c_{2}^{2}
&\leq&\frac{1}{4}\{[2ab(1+t^{2})+t]-2\sqrt{D(a,b,t)}\}
\end{eqnarray}
which agree with the P-representation condition (51). Any separable state, in particular, the P-representable state satisfies the condition (38) by its construction. The relations (51) and (53) now show that  any Gaussian state which satisfies the condition (38) satisfies the P-representation, and thus (38) gives  a necessary and sufficient criterion for the separability of two-party Gaussian systems \cite{simon}.

We now come to the weaker form of the separability condition (40). It is confirmed that the condition (40) when applied to the covariance matrix $V$ in (41) gives rise to (see also the second condition in (52))  
\begin{eqnarray}
\sqrt{[ar_{1}+\frac{a}{r_{1}}-1][br_{2}+\frac{b}{r_{2}}-1]}
\geq \sqrt{r_{1}r_{2}}|c_{1}|+\frac{|c_{2}|}{\sqrt{r_{1}r_{2}}}.
\end{eqnarray}
If one imposes the condition (46),
the left-hand side of (54) becomes 
\begin{eqnarray}
&&\sqrt{[ar_{1}+\frac{a}{r_{1}}-1][br_{2}+\frac{b}{r_{2}}-1]}
\\
&=&\sqrt{(ar_{1}-1/2)(br_{2}-1/2)} + \sqrt{(a/r_{1}-1/2)(b/r_{2}-1/2)}\nonumber.
\end{eqnarray}
See Appendix C. 
The weaker form of separability condition (40) together with (46) thus gives 
\begin{eqnarray}
&&\sqrt{(ar_{1}-1/2)(br_{2}-1/2)} + \sqrt{(a/r_{1}-1/2)(b/r_{2}-1/2)}\nonumber\\
&&\geq \sqrt{r_{1}r_{2}}|c_{1}|+\frac{|c_{2}|}{\sqrt{r_{1}r_{2}}}
\end{eqnarray} 
which is confirmed to correspond to the separability condition of Duan et al., eq. (16) in \cite{duan}, when converted into their notation. 

When one sets $|c_{2}|=t|c_{1}|$ in (56), one obtains the condition
\begin{eqnarray}
&&\sqrt{(ar_{1}-1/2)(br_{2}-1/2)} + \sqrt{(a/r_{1}-1/2)(b/r_{2}-1/2)}\nonumber\\
&&\geq [\sqrt{r_{1}r_{2}}+\frac{t}{\sqrt{r_{1}r_{2}}}]|c_{1}|.
\end{eqnarray} 
One can confirm that this relation (57) when regarded as a bound on
$|c_{1}|$ agrees with the P-representation condition (51) if one uses the first equality in (51).
We thus conclude that the weaker separability condition (40)
when combined with the squeezing defined by (48) provides a necessary and sufficient criterion for the separability of two-party Gaussian states. This proof, which was sketched in 
\cite{fujikawa}, is much simpler and more explicit than the original proof in \cite{duan} which was also completed in \cite{fujikawa}.

In the above analysis, we implicitly used the solution of (46)
\begin{eqnarray}
r_{2}(r_{1})=\frac{4b}{[\sqrt{(1-X)^{2}+16b^{2}X}+(1-X)]}
\end{eqnarray}
with $X(r_{1})=(2a/r_{1}-1)/(2ar_{1}-1)$
which assumes $X(1)=1$ and $X(2a)=0$, and 
thus $r_{2}(1)=1$ and $r_{2}(2a)=2b$.

\section{Conclusion}

We have derived a 
separability criterion for continuous variable two-party
systems without referring to the non-negativity of the partially transposed density matrix. This condition on the second moment (or covariance matrix) is generally more stringent than the condition of Simon \cite{simon} which is based the non-negativity of the partially transposed density matrix.
Our criterion may thus be useful as a {\em necessary} condition for
the separability of general two-party systems. We have also shown that the separability condition used by Duan et al. \cite{duan} is generally weaker than the separability condition of Simon. 
Nevertheless, at the boundary of the P-representation condition for Gaussian two-party systems, all these separability conditions
become equivalent to each other if one uses a suitable squeezing 
operation.
This fact is most easily shown by using the explicit analytic formulas of squeezing parameters which establish the equivalence of the  $Sp(2,R)\otimes Sp(2,R)$ invariant separability condition with the P-representation condition of Gaussian states
\cite{fujikawa}. 
\\

I thank K. Shiokawa for an informative discussion at the very beginning of the present study.    

\appendix

\section{Standard form of $V$}

We recall the elements 
of $Sp(2,R)$ 
\begin{eqnarray}
S=\left(\begin{array}{cc}
  \cos\theta&\sin\theta\\
  -\sin\theta&\cos\theta\\
            \end{array}\right), \ \ \ \ 
S=\left(\begin{array}{cc}
  x&0\\
  0&\frac{1}{x}\\
            \end{array}\right)
\end{eqnarray}
which satisfy $SJS^{T}=J$.  One can bring $V$ in (17) to the 
standard form 
\begin{eqnarray}
V=\left(\begin{array}{cccc}
  a&0&c_{1}&0\\
  0&a&0&c_{2}\\
  c_{1}&0&b&0\\
  0&c_{2}&0&b\\            
  \end{array}\right)
\end{eqnarray}
by suitable $Sp(2,R)\otimes Sp(2,R)$ 
transformations~\cite{duan,
simon}; real
symmetric $A$ and $B$ can be made diagonal by two-dimensional 
rotations with suitable parameters $\theta$ in (A1) and 
then applying the second elements in (A1) with suitable 
parameters $x$, $A$ and $B$ are 
made proportional to the unit matrix. After these 
transformations $C$ remains real. By applying a suitable 
two-dimensional orthogonal
transformation $S_{1}\otimes S_{2}$, which is an element of 
$Sp(2,R)\otimes Sp(2,R)$, we can diagonalize $C$ 
\begin{eqnarray}
S_{1}CS^{T}_{2}=\left(\begin{array}{cc}
  c_{1}&0\\
  0&c_{2}\\            
  \end{array}\right).
\end{eqnarray}
By this way we arrive at (A2).

\section{P-representation}
 
We define the generating function of all the correlations
(or moments) of  dynamical variables by
\begin{eqnarray}
\chi(\lambda, \eta)={\rm Tr}(\hat{\rho}\exp\{
i(\lambda_{1}\hat{q}_{1}+
\lambda_{2}\hat{p}_{1}+
\eta_{1}\hat{q}_{2}+\eta_{2}\hat{p}_{2})\})
\end{eqnarray}
where $\lambda_{1}\sim \eta_{2}$ are real numbers. By expanding 
$\chi(\lambda, \eta)$ in powers of $\lambda_{1}\sim \eta_{2}$, one can generate all the moments of dynamical variables. One may generally choose 
$\langle \hat{q}_{1}\rangle=\langle\hat{p}_{1}\rangle=
\langle \hat{q}_{2}\rangle=\langle\hat{p}_{2}\rangle=0$.
Following the convention in this field, we define the Gaussian
states by 
\begin{eqnarray}
\chi(\lambda, \eta)=
\exp\{-\frac{1}{2}(\lambda_{1},\lambda_{2},\eta_{1},\eta_{2})V(\lambda_{1},\lambda_{2},\eta_{1},\eta_{2})^{T}\}
\end{eqnarray}
where $V$ is the covariance matrix in (17), namely, all the 
correlation functions are determined by the second moments.
Note that c-numbers $\lambda_{1}\sim \eta_{2}$ are commuting and thus all the operators are automatically symmetrized in (B1). One can write (B1) as 
\begin{eqnarray}
\chi(\lambda, \eta)={\rm Tr}(\hat{\rho}\exp\{i(
\lambda^{\star}\hat{a}+\lambda\hat{a}^{\dagger}
+\eta^{\star}\hat{b}+\eta\hat{b}^{\dagger})\})
\end{eqnarray}
with
\begin{eqnarray}
&&\hat{a}=\frac{1}{\sqrt{2}}(\hat{q}_{1}+i\hat{p}_{1}),\ \ 
\hat{b}=\frac{1}{\sqrt{2}}(\hat{q}_{2}+i\hat{p}_{2}),\nonumber\\
&&\lambda=\frac{1}{\sqrt{2}}(\lambda_{1}+i\lambda_{2}),\ \ 
\eta=\frac{1}{\sqrt{2}}(\eta_{1}+i\eta_{2}).
\end{eqnarray}
The Gaussian state is called P-representable if the density matrix is written as
\begin{eqnarray}
\hat{\rho}&=&\int d^{2}\alpha\int d^{2}\beta P(\alpha,\beta)
|\alpha,\beta\rangle\langle\alpha,\beta|
\end{eqnarray}
where $|\alpha,\beta\rangle$ is the (over complete) coherent state defined by
\begin{eqnarray}
&&\hat{a}|\alpha,\beta\rangle=\alpha|\alpha,\beta\rangle,\ \ 
\hat{b}|\alpha,\beta\rangle=\beta|\alpha,\beta\rangle,\ \
\langle\alpha,\beta|\alpha,\beta\rangle=1\nonumber\\
\end{eqnarray}
or to be explicit
\begin{eqnarray}
|\alpha,\beta\rangle=e^{\alpha\hat{a}^{\dagger}
-\frac{1}{2}|\alpha|^{2}}|0\rangle\otimes 
e^{\beta\hat{b}^{\dagger}-\frac{1}{2}|\beta|^{2}}|0\rangle.
\end{eqnarray}
Thus the P-representable states are separable.

By using the density matrix (B5) in (B3) and after normal 
ordering the exponential factor in (B3), we have
\begin{eqnarray}
\chi(\lambda, \eta)&=&\int d^{2}\alpha\int d^{2}\beta P(\alpha,\beta)\nonumber\\
&&\times \exp\{i(
\lambda^{\star}\alpha+\lambda\alpha^{\star}
+\eta^{\star}\beta+\eta\beta^{\star})\}\nonumber\\
&&\times \exp\{-\frac{1}{2}(
|\lambda|^{2}+|\eta|^{2})\}
\end{eqnarray}
or, if one combines this expression with (B2) we have
\begin{eqnarray}
&&\exp\{-\frac{1}{2}(\lambda_{1},\lambda_{2},\eta_{1},\eta_{2})
(V-\frac{1}{2}I)(\lambda_{1},\lambda_{2},\eta_{1},\eta_{2})^{T}\}
\nonumber\\
&=&\int d^{2}\alpha\int d^{2}\beta P(\alpha,\beta)\nonumber\\
&&\times \exp\{i(
\lambda_{1}\alpha_{1}+\lambda_{2}\alpha_{2}
+\eta_{1}\beta_{1}+\eta_{2}\beta_{2})\}\nonumber\\
&=&\int d^{2}\alpha\int d^{2}\beta P(\alpha,\beta)\nonumber\\
&&\times \exp\{i(
\lambda^{\star}\alpha+\lambda\alpha^{\star}
+\eta^{\star}\beta+\eta\beta^{\star})\}
\end{eqnarray}
with $\alpha=(\alpha_{1}+i\alpha_{2})/\sqrt{2}$ and 
$\beta=(\beta_{1}+i\beta_{2})/\sqrt{2}$. Thus 
$P(\alpha,\beta)$ in (B9) is given by
\begin{eqnarray}
P(\alpha,\beta)&=&\frac{\sqrt{det P}}{4\pi^{2}}\\
&&\times\exp\{-\frac{1}{2}(\alpha_{1},\alpha_{2},\beta_{1},\beta_{2})P(\alpha_{1},\alpha_{2},\beta_{1},\beta_{2})^{T}\}
\nonumber
\end{eqnarray}
with a matrix $P$
\begin{eqnarray}
P^{-1}=V-\frac{1}{2}I\geq 0,
\end{eqnarray}
which defines the condition for the existence of the well-defined P-representation.

The P-representation is characterized by the weight factor
$P(\alpha,\beta)$ which is in turn determined by the matrix $P$.
We now notice that the right-hand side of the formula (B9), when expanded in terms of $\lambda$'s and $\eta$'s, generates the correlations of the form
\begin{eqnarray}
&&\int d^{2}\alpha\int d^{2}\beta P(\alpha,\beta)
\alpha^{\star}\alpha\\
&&=\int d^{2}\alpha\int d^{2}\beta P(\alpha,\beta)
\langle\alpha,\beta|\hat{a}^{\dagger}
|\alpha,\beta\rangle \langle\alpha,\beta|\hat{a}
|\alpha,\beta\rangle,\nonumber
\end{eqnarray}
for example. This may be compared to (20). By recalling 
(B5), this (B12) shows that all
the second moments on the right-hand side of (B9), which are 
determined by $P$, coincide with $\tilde{V}$ in (20) (if one chooses 
$\langle \hat{a}^{\dagger}\rangle=\langle\hat{a}\rangle=
\langle \hat{b}^{\dagger}\rangle=\langle\hat{b}\rangle=0$).
This property establishes the special relation (33) of the 
P-representation, namely, $\tilde{V}$ is determined by
\begin{eqnarray}
\tilde{V}=P^{-1}.
\end{eqnarray}

\section{Proof of (55)}

One can show
\begin{eqnarray}
&&\sqrt{[(n_{1}+n_{2})-1][(m_{1}+m_{2})-1]}\nonumber\\
&&\geq \sqrt{[n_{1}-\frac{1}{2}][m_{1}-\frac{1}{2}]}+ \sqrt{[n_{2}-\frac{1}{2}][m_{2}-\frac{1}{2}]}
\end{eqnarray}
where the equality holds only when the condition 
\begin{eqnarray}
\frac{(n_{2}-\frac{1}{2})}{(n_{1}-\frac{1}{2})}=\frac{(m_{2}-\frac{1}{2})}{(m_{1}-\frac{1}{2})}
\end{eqnarray}
is satisfied. This relation (C1) is established by considering 
\begin{eqnarray}
f(x)=\sqrt{[n_{1}+x(n_{2}-n_{1})-\frac{1}{2}]
[m_{1}+x(m_{2}-m_{1})-\frac{1}{2}]}\nonumber\\
\end{eqnarray}
and
\begin{eqnarray}
f^{''}(x)&=&-\frac{1}{4}[(n_{1}-\frac{1}{2})(m_{2}-\frac{1}{2})
-(n_{2}-\frac{1}{2})(m_{1}-\frac{1}{2})]^{2}\nonumber\\
&\times&
[m_{1}+x(m_{2}-m_{1})-\frac{1}{2}]^{-3/2}\nonumber\\
&\times&
[n_{1}+x(n_{2}-n_{1})-\frac{1}{2}]^{-3/2}< 0
\end{eqnarray}
except for (C2) for which $f^{''}(x)=0$. By using the property of the convex function $2f(1/2)\geq f(1)+f(0)$
one establishes (C1). If one sets $n_{1}=ar_{1}, \ n_{2}=a/r_{1}, \ m_{1}=br_{2}$ and $ m_{2}=b/r_{2}$, (C1) gives (55) while (C2) corresponds to (46).

\end{document}